\documentclass[10pt,twocolumn,letterpaper]{article}

\usepackage[pagenumbers]{cvpr} 
\usepackage{multirow}
\usepackage{graphicx}
\usepackage{multicol}
\usepackage{makecell}
\usepackage{booktabs} 
\usepackage{colortbl}
\usepackage{wasysym}

\usepackage[dvipsnames]{xcolor}

\definecolor{cvprblue}{rgb}{0.21,0.49,0.74}
\usepackage[pagebackref,breaklinks,colorlinks,citecolor=cvprblue]{hyperref}

\def\paperID{12345}
\def\confName{CVPR}
\def\confYear{2024}

\title{Multimodal Infusion Tuning for Large Models}

\author{Hao Sun\\
Zhejiang University\\
Hangzhou, China\\
{\tt\small sunhaoxx@zju.edu.cn}
\and
Yu Song\\
Ritsumeikan University\\
Shiga, Japan\\
{\tt\small yusong@fc.ritsumei.ac.jp}
\and
Xinyao Yu\\
Zhejiang University\\
Hangzhou, China\\
{\tt\small xinyaoyu@zju.edu.cn}
\and
Jiaqing Liu\\
Ritsumeikan University\\
Shiga, Japan\\
{\tt\small liu-j@fc.ritsumei.ac.jp}
\and
Yen-Wei Chen\\
Ritsumeikan University\\
Shiga, Japan\\
{\tt\small chen@is.ritsumei.ac.jp}
\and
Lanfen Lin\\
Zhejiang University\\
Hangzhou, China\\
{\tt\small llf@zju.edu.cn}
}

\begin{document}
\maketitle
\begin{abstract}
Recent advancements in large-scale models have showcased remarkable generalization capabilities in various tasks. However, integrating multimodal processing into these models presents a significant challenge, as it often comes with a high computational burden. To address this challenge, we introduce a new parameter-efficient multimodal tuning strategy for large models in this paper, referred to as Multimodal Infusion Tuning (MiT). MiT leverages decoupled self-attention mechanisms within large language models to effectively integrate information from diverse modalities such as images and acoustics. In MiT, we also design a novel adaptive rescaling strategy at the attention head level, which optimizes the representation of infused multimodal features. Notably, all foundation models are kept frozen during the tuning process to reduce the computational burden and only 2.5\% parameters are tunable. We conduct experiments across a range of multimodal tasks, including image-related tasks like referring segmentation and non-image tasks such as sentiment analysis. Our results showcase that MiT achieves state-of-the-art performance in multimodal understanding while significantly reducing computational overhead(10\% of previous methods). Moreover, our tuned model exhibits robust reasoning abilities even in complex scenarios.
\end{abstract}

\begin{figure*}
    \centering
    \begin{minipage}{0.95\linewidth}
        \centering
        \includegraphics[width=1.0\textwidth]{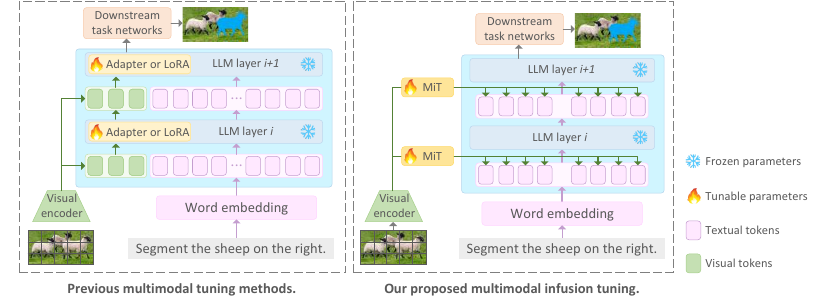}
    \end{minipage}
  \caption{The comparison between previous multimodal tuning methods(\textit{left})~\cite{ren2023pixellm} and our proposed multimodal infusion tuning(\textit{right}). The referring segmentation task is taken as an example. Previous methods usually treat visual embeddings as tokens, which are prefixed to textual tokens and some PEFT adapters are employed to tune the LLMs on downstream tasks. In our proposed multimodal infusion tuning, we infuse the visual information into each textual token in a linear manner, which is more fine-grained and dose not introduce extra tokens to pretrained LLMs.}
  \label{fig:overview}
\end{figure*}

\section{Introduction}
Large language models (LLMs), benefiting from their immense scale in model sizes and training data, have demonstrated impressive capabilities across a wide range of real-world tasks. The prominence of LLMs is evident in the widespread attention they have garnered from both academia and industry, with notable examples being OPT~\cite{zhang2022opt} and LLaMA~\cite{touvron2023llama}. However, training a new large model is highly resource-intensive, leading to the development of numerous parameter-efficient fine-tuning (PEFT) methods, such as LLaMA-adapter~\cite{zhang2023llama} and (IA)$^3$~\cite{liu2022few}.

Currently, the majority of PEFT approaches are tailored for text-only data~\cite{li2021prefix,liu2022few}, posing a challenge in handling multimodal information. Empowering LLMs with multimodal data, such as images and acoustics, remains a significant challenge. To this end, some researchers have proposed integrating image information into LLMs through PEFT, exemplified by approaches like FROMAGe~\cite{koh2023grounding} and mPLUG-Owl~\cite{ye2023mplug}. Most of these methods tend to concatenate or prefix the vision embedding with text tokens and feed them directly to the LLMs, as shown in Figure~\ref{fig:overview}(\textit{left}). However, we identify two drawbacks in this approach. Firstly, the concatenation of tokens significantly increases memory consumption, growing quadratically due to the self-attention mechanism's $L^{2}$ memory requirement (where $L$ represents the length of tokens). Secondly, direct prefix approaches are deemed coarse-grained, leading to insufficient interactions among representations from multiple modalities.

To address these issues, we propose a new tuning framework termed multimodal infusion tuning (MiT). The whole tuning procedure is designed to be linear, ensuring low memory consumption. This infusion strategy allows us to endow LLMs with the capability to process multimodal information. One major difference between MiT and other methods is that we decouple self-attention in-depth, and inject multimodal information at a more fine-grained aspect(shown in Figure~\ref{fig:overview}(\textit{right})). Specifically, we decouple the self-attention of large language models and progressively infuse the global representations of other modalities into the keys and values of text embedding. Additionally, we introduce a novel adaptive rescaling strategy at the attention head level to enhance collaboration between different modalities. We freeze the parameters in pretrained models and exclusively fine-tune those newly introduced in MiT(around only 2.5\% of all parameters). Another distinguishing feature of our approach compared to previous methods is its versatility. MiT is not limited to processing image data; it can also effectively handle other types of data, including acoustic and facial features. For various multimodal tasks, we employ different downstream networks during tuning, such as a lightweight decoder for referential segmentation and a fully connected layer for sentiment analysis. 

We conduct experiments on seven datasets encompassing three tasks(referring segmentation, image-text classification, and sentiment analysis), and the results demonstrate that MiT achieves state-of-the-art performance while maintaining efficiency(our proposed method requires only 0.47 TFLOPs). Moreover, leveraging the powerful language modeling capabilities of LLMs, our MiT is capable of performing reasoning and segmentation tasks guided by complex texts. For instance, the model can correctly segment phrases like \textit{spicy vegetables} instead of just \textit{peppers}, or \textit{best friend of human} instead of simply \textit{dog}.

In summary, our paper presents four key contributions:
\begin{itemize}
\item We introduce a fine-grained multimodal tuning strategy for LLMs, named MiT. This strategy is designed to efficiently accommodate various types of multimodal data, allowing for their progressive infusion into LLMs.
\item We develop an adaptive rescaling strategy at the head level, facilitating the collaboration of information from different modalities to enhance interaction.
\item Our tuned model demonstrates sophisticated reasoning capabilities in complex language scenarios. This suggests that the MiT approach can be effectively applied to a wider range of complex tasks and scenarios, representing the effectiveness of our method.
\item Our approach achieves state-of-the-art performance on seven evaluated datasets, where the pretrained LLM is kept frozen during training and only around 2.5\% all parameters are trainable.
\end{itemize}

\section{Related Works}
Our work builds upon existing LLM and multimodal fine-tuning methods. In this section, we will introduce some previous methods relevant to our work.

\subsection{Large Language Models and Tuning}
In recent years, large language models (LLMs) have demonstrated significant advancements in in-context learning and long-term generation, as showcased by models such as ChatGPT, OPT~\cite{zhang2022opt}, and LLaMA~\cite{touvron2023llama}. These models have significantly improved the capabilities of language understanding and generation. To further enhance the application of LLMs in downstream tasks, various methods have been introduced. For example, adapter-tuning~\cite{houlsby2019parameter} and prefix-tuning ~\cite{li2021prefix} are two prominent techniques that focus on fine-tuning the model while preserving the original parameters. This approach offers two key benefits: it reduces computational overhead and maintains the proficiency of the LLM trained on large corpora.

Recently, there has been a surge in novel Prompt Encoding for Fine-Tuning (PEFT) methods, such as LLaMA-adapter~\cite{zhang2023llama} and (IA)$^{3}$~\cite{liu2022few}. Notably, (IA)$^{3}$ introduces learnable vectors that are multiplied with the keys (K) and values (V) in self-attention mechanisms, effectively reducing memory usage. Our proposed method draws considerable inspiration from (IA)$^{3}$, although it is specifically tailored for multimodal data and tasks. This adaptation allows our method to leverage the strengths of (IA)$^{3}$ while addressing the unique challenges posed by multimodal learning.

\begin{figure*}
    \centering
    \begin{minipage}{1.0\linewidth}
        \centering
        \includegraphics[width=1.0\textwidth]{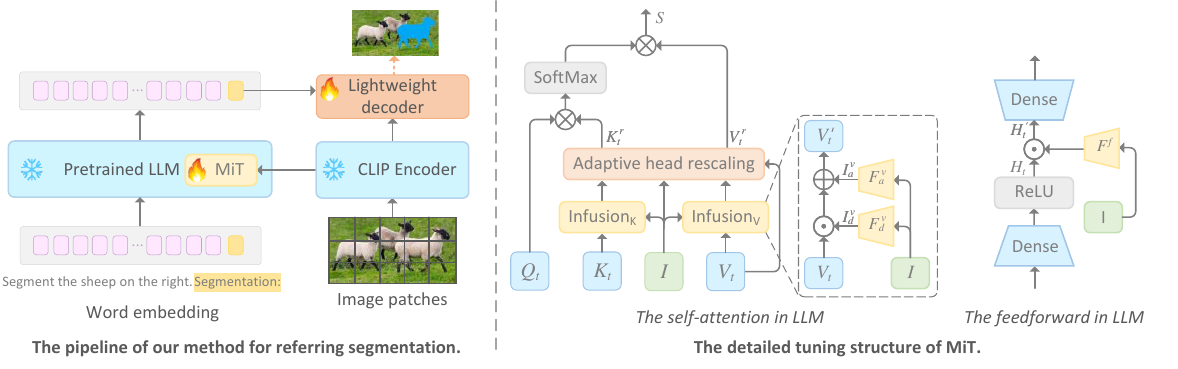}
    \end{minipage}
  \caption{The pipeline of our proposed method(\textit{left}) and the detailed structure of MiT(\textit{right}). The referring segmentation is shown as the example, which is part of a broader framework that includes several other tasks, such as image-text classification and sentiment analysis. In MiT, we infuse the visual information in both self-attention and feedforward module. The procedure is kept linear for computational burden consideration.}
  \label{fig:mit}
\end{figure*}

\subsection{Multimodal Tuning of LLMs}
While LLMs have demonstrated remarkable success in natural language processing, their ability to perceive and understand other modalities, such as images and acoustics, remains a challenge. Recent efforts have aimed to equip LLMs with multimodal perception capabilities. For instance, Flamingo~\cite{alayrac2022flamingo} employs a frozen image encoder and integrates multimodal signals using gated cross-attention, showcasing the potential of LLMs for understanding multiple modalities. Building upon this, BLIP-2~\cite{li2023blip} introduces the Q-Former module, a sophisticated structure designed to align image and text embeddings by extracting common semantics between modalities. This approach has found widespread adoption in various studies due to its effectiveness. Furthermore, Driess et al.~\cite{driess2023palm} proposed PaLM-E, which directly incorporates multimodal information as input and has exhibited significant efficacy across numerous tasks. Similarly, FROMAGe~\cite{koh2023grounding} introduces linear transformations to ground text features in the visual domain, facilitating seamless translation between texts and images. With the release of LLaMA~\cite{touvron2023llama}, efforts have been made to adapt the model for multimodal tasks, as demonstrated by LLaMA-adapter~\cite{zhang2023llama}. While initially focused on text-based tasks, LLaMA-adapter has shown promising multimodal performance by concatenating image embeddings with text tokens. However, this approach leads to quadratic increases in memory consumption due to the attention mechanism. In contrast, our method also equips LLMs with multimodal understanding capabilities but employs a novel infusion tuning strategy, resulting in linear memory consumption. Moreover, our approach is versatile, capable of processing not only image data but also incorporating information from other modalities, such as acoustic and facial features, particularly beneficial in sentiment analysis.

\section{Multimodal Infusion Tuning}
Our method seamlessly integrates multimodal information into a pretrained LLM while keeping its parameters frozen. The overall pipeline is illustrated in Figure~\ref{fig:mit}(\textit{left}). We introduce a tunable MiT module into the pretrained LLM, facilitating the integration of representations from other modalities. Through tuning, the LLM progressively acquires the capability to process multimodal signals as MiT module is zero-initialized. The resulting interacted textual features, combined with visual features in tasks involving images, are then fed into various downstream networks for predictions. Consequently, our method finds applications across diverse domains, including referring segmentation, image-text classification, and sentiment analysis(Figure~\ref{fig:mit} shows the example of referring segmentation).

\subsection{Architecture Design}
As illustrated in Figure~\ref{fig:mit}(\textit{right}), we infuse the multimodal information in both self-attention and feed-forward module of pretrained LLMs. We design MiT as a plugable module, so we can leverage the capabilities of the LLM learnt from large scale corpus. 

Without loss of generality, assume that we have an global image representation extracted by CLIP image encoder~\cite{radford2021learning}: $I\in R^{d_I}$, where $d_I$ is the corresponding feature dimension. In textual self-attention of LLM, the text embeddings are projected into query($Q_{t}$), key($K_{t}$), and value($V_{t}$). We delve into textual self-attention and integrate $I$ into $K_{t}$ and $V_{t}$. Take $V_t$ for example, to fit it into the space of text representations, we employ two transformations:
\begin{equation}
\begin{aligned}
I_{d}^{v} &= F^{v}_{d}(I) = I\cdot W^{v}_{d} + b^{v}_{d},\\
I_{a}^{v} &= F^{v}_{a}(I) = I\cdot W^{v}_{a} + b^{v}_{a},\\
\label{equ:image_trans}
\end{aligned}
\end{equation}
where $W^{v}_{d},W^{v}_{a}\in R^{d_{I}\times d_{T}}$, $b^{v}_{d},b^{v}_{a}\in R^{d_{T}}$ are learnable matrices and vectors. In order to keep the computational burden linear, we infuse the transformed representation into $V_{t}$ via element-wise multiplication and addition(the Infusion$_{V}$ in Figure~\ref{fig:mit}):
\begin{equation}
V_t^{'} = V_t \cdot I_{d}^{v} + I_{a}^{v}.
\end{equation}
The same is true the infusion of $K_t$, which can be represented as:
\begin{equation}
K_t^{'} = K_t \cdot I_{d}^{k} + I_{a}^{k},
\end{equation}
where $I_{d}^{k}$ and $I_{a}^{k}$ are generated in the same approach as Equation~\ref{equ:image_trans}(Infusion$_{K}$ in Figure~\ref{fig:mit}).

Numerical instabilities may emerge when integrating multimodal representations into LLM due to the modality gap. To address this issue, we have implemented an adaptive rescaling mechanism at the head-level. Given that current self-attention mechanisms operate with multiple heads, the tensors $V_t^{'}$ and $K_t^{'}$ are reshaped as $R^{h\times d_h}$ for each token prior to computing the attention map, where $h$ represents the number of heads, and $d_{h}=d_{T}/h$ denotes the dimensionality per head. Consequently, we have introduced a learnable rescaling vector $L\in R^{h}$, which is combined with the similarity between the text value $V_{t}$ and the image embedding $I$ at the head level:
\begin{equation}
\begin{aligned}
L^{'} = L + \frac{V_{t}\cdot I}{||V_t||\cdot||I||}.
\end{aligned}
\end{equation}
Then we element-wise multiply the normalized rescaling vector to the infused value and key:
\begin{equation}
\begin{aligned}
V^{r}_{t} = V^{'}_{T} \cdot \sigma(L^{'}),\\
K^{r}_{t} = K^{'}_{T} \cdot \sigma(L^{'}),\\
\end{aligned}
\end{equation}
where $\sigma$ is the sigmoid function. The rescaled value and key, alongside with query($Q_T$), are then employed to perform the self-attention:
\begin{equation}
\begin{aligned}
S = \mathrm{softmax}(Q_t K^{rT}_{t}/\sqrt{d_{T}}) V^{r}_{t}.
\end{aligned}
\end{equation}
Finally after the self-attention, the integration is also performed in feed-forward module:
\begin{equation}
\begin{aligned}
H_{t}^{'} &= H_{t}\cdot F^{f}(I) = H_{t}\cdot(I\cdot W^{f} + b^{f}),\\
\end{aligned}
\end{equation}
where $W^{f}\in R^{d_{I}\times d_{T'}}$, $b^{f}\in R^{d_{T'}}$, $H_{t}\in R^{d_{T'}}$ is the activated textual hidden embedding in LLM, and $d_{T'}$ is the corresponding hidden size.

In this approach, we can seamlessly integrate multimodal information with a linear computational burden, yet in a more fine-grained manner. Notably, all learnable vectors to be added are initialized to 0 (e.g., $I_{d}^{v}$), and all vectors to be multiplied are initialized to 1 (e.g., $I_{a}^{v}$) in MiT. Therefore, the LLMs can progressively acquire the ability for multimodal understanding with minimal computational load.

\begin{figure*}[ht]
    \centering
    \begin{minipage}{1.0\linewidth}
        \centering
        \includegraphics[width=0.7\textwidth]{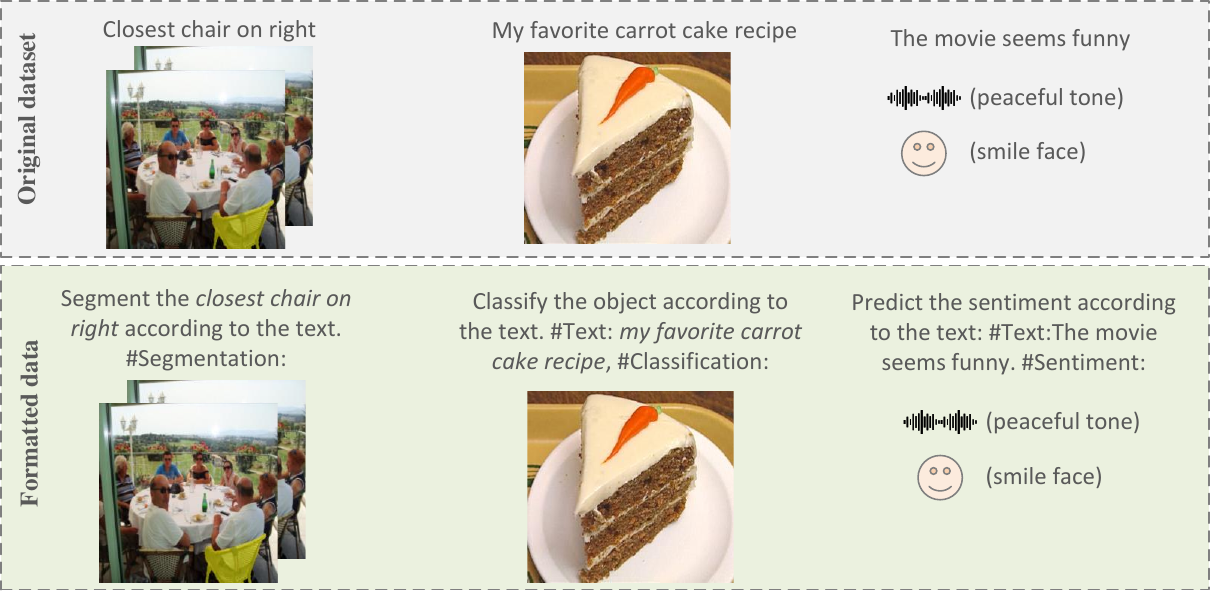}
    \end{minipage}
  \caption{The formation of our employed dataset and tasks: referring segmentation(\textit{left}), image-text classification(\textit{middle}), and sentiment analysis(\textit{right}). For different dataset and tasks, we design different templates, so as to excavate the capabilities of LLMs obtained in pretraining.}
  \label{fig:data}
\end{figure*}

\subsection{Downstream Task Tuning}
We fine-tune our MiT on three tasks for multimodal understanding: referring segmentation, image-text classification, and multimodal sentiment analysis. Among these tasks, sentiment analysis does not involve images, as it focuses on analyzing text or other forms of non-image data (for privacy considerations, only processed facial features are provided in the datasets evaluated). In contrast, the other two tasks, object detection and image classification, require image data for analysis. For tasks involving images, we utilize a frozen CLIP encoder to generate global semantic representations and features of images at different levels, as illustrated in Figure~\ref{fig:mit}. Alternatively, for tasks not involving images, we employ a trainable transformer feature encoder to extract multimodal embeddings. Since the LLMs employed in our approach are decoder-only structures, we extract the semantic representations for downstream tasks from the last token of the entire sequences.

For referring segmentation, the goal is to segment the object from the image according to a descriptive. Under this scenario, we design a light-weight decoder, which takes infused text embedding and multi-level image features as input, and generate the predicted mask. The DICE loss~\cite{milletari2016v} is employed in our framework:
\begin{equation}
\begin{aligned}
\mathcal{L}_{seg} = \mathcal{L}_{DICE} = 1- \frac{2|\hat{y}\cap y_{gt}|}{|\hat{y}|+ |y_{gt}|}
\end{aligned}
\end{equation}
where $\hat{y}$ is the predicted mask and $y_{gt}$ is the ground truth. For image-text classification, the global image representation and infused text embedding are concatenated, followed by a prediction head. Like general approaches, the cross-entropy loss is employed for this task:
\begin{equation}
\begin{aligned}
\mathcal{L}_{cls} = \mathcal{L}_{CE} = -\sum_{i=1}^{n}y_{gt,i}\log (\hat{y}_{i})
\end{aligned}
\end{equation}
where $n$ is the number of training samples in a mini-batch.
For multimodal sentiment analysis, we only utilize the infused text embedding for prediction, as text is dominant in this task(shown in Figure~\ref{fig:mit}). As sentiment analysis is a regression task(sentiment tendency), we employ the RMSE as the loss function following previous works~\cite{hazarika2020misa,han2021bi}:
\begin{equation}
\begin{aligned}
\mathcal{L}_{msa} = \mathcal{L}_{RMSE} = \sqrt{\frac{1}{n}\sum_{i=1}^{n}(\hat{y}_{i}-y_{gt,i})}.
\end{aligned}
\end{equation}

\subsection{Implementation Details}
We evaluate our method on RefCOCO~\cite{yu2016modeling}, RefCOCO+~\cite{yu2016modeling}, and G-Ref~\cite{mao2016generation} for referring segmentation. There are 19994, 19992, and 26711 images, with
50000, 49856, and 54822 references, respectively. For image-text classification, we employ two datasets: UPMC-Food101~\cite{wang2015recipe} and SNLI-VE~\cite{xie2018visual}, which contain 90840 and 565286 image-text pairs, respectively. UPMC-Food101 provides 101-category classification task while SNLI-VE aims to reason the semantic relationship(entailment, neutral, or contradiction) between text hypothesis and image premise. As for the sentiment analysis, we use two popular dataset: MOSI~\cite{zadeh2016multimodal} and MOSEI~\cite{zadeh2018multimodal}, in which the sentiment tendency are annotated within [-3, +3]. The goal is to predict the sentiment tendency based on corresponding texts, acoustic and facial features.

We use the popular LLaMA~\cite{touvron2023llama} with 7B parameters as the textual foundation model. For image branch, we employ the CLIP ViT-L-336 model~\cite{radford2021learning} to generate the global representations. In MiT, we infuse the multimodal representation into last certain layers in LLM, and the detailed effect are discussed in Section~\ref{sec:abl}. There are three layers in the light-weight decoder for referring segmentation to maintain efficiency. In sentiment analysis task, three-layer transformer blocks are employed to encoder acoustic and facial features. Inherited from pretrained models, $d_{I}$, $d_{T}$, and $d_{T'}$ are set to 768, 4096, and 11008, respectively. 

Our models are built under the PyTorch framework and tuned with half-precision(bfloat16~\cite{abadi2016tensorflow}). As only 2.5\% parameters are tunable and no extra memory are required by self-attention, MiT is memory and compute efficient. We use the Adam optimizer and set the learning rate to 0.00004. We train the model for 30 epochs and the learning rate is decayed by 0.1 every 10 epochs. Our experiments are conducted on two NVIDIA RTX 4090 GPUs. The batch size is set to 8 on each GPU. The whole tuning process takes 20 hours on RefCOCO and 1 hours on MOSI.

\section{Experimental Results}
To better utilize the capabilities of LLM obtained from pretraining, we format the original inputs in the evaluated dataset, as shown in Figure~\ref{fig:data}. For example, we format the input sentence as \textit{Segment the \{description\} according to the text. \#Segmentation:} for referring segmentation.  We use overall Intersection-over-Unions(oIoU) as metrics for referring segmentation following previous works. Accuracy and F1-score are utilized as metrics for classification. For sentiment analysis, we use mean absolute error(MAE), Pearson correlation coefficient(Corr) and corresponding accuracy as metrics.

\begin{table*}[]
\centering
\caption{The results of our method on referring segmentation benchmarks: RefCOCO, RefCOCO+, and RefCOCOg. The metrics in the table is oIoU. LLM/TFLOPs means the LLM employed and the methods' overall computation requirement. RefCOCOg(U) means the UMD partition of RefCOCOg dataset.}
\label{Tab:resultRef}
\begin{tabular}{c|c|ccc|ccc|cc}
\toprule 
\multirow{2}{*}{Method} & \multirow{2}{*}{LLM/TFLOPs} & \multicolumn{3}{c|}{RefCOCO} & \multicolumn{3}{c|}{RefCOCO+} & \multicolumn{2}{c}{RefCOCOg(U)} \\
 &  & val & testA & testB & val & testA & testB & val & test \\
\midrule
CRIS~\cite{wang2022cris} & - & 70.5 & 73.2 & 66.1 & 62.3 & 68.1 & 53.7 & 59.9 & 60.4 \\
LAVT~\cite{yang2022lavt} & - & 72.7 & 75.8 & 68.8 & 62.1 & 68.4 & 55.1 & 61.2 & 62.1 \\
ReLA~\cite{liu2023gres} & - & 73.8 & 76.5 & 70.2 & 66.0 & 71.0 & 57.7 & 65.0 & 66.0 \\
SEEM~\cite{zou2023segment} & - & - & - & - & - & - & - & 64.6 & - \\
\midrule
LISA~\cite{lai2023lisa} & LLaMA2-13B/10.24 & \textbf{74.1} & 76.5 & \textbf{71.1} & 62.4 & 67.4 & 56.5 & 66.4 & \textbf{68.5} \\
PixelLM~\cite{ren2023pixellm} & LLaMA2-13B/6.65 & 73.0 & 76.5 & 68.2 & 66.3 & 71.7 & \textbf{58.3} & \textbf{69.3} & 70.5 \\
MiT(ours) & LLaMA2-7B/\textbf{0.47} & 73.1 & \textbf{77.1} & 68.4 & \textbf{66.9} & \textbf{72.2} & 57.6 & 65.4 & \textbf{68.5} \\
\bottomrule
\end{tabular}
\end{table*}

\subsection{Results on Referring Segmentation}
Table~\ref{Tab:resultRef} shows the results on referring segmentation benchmarks. We compare our MiT with previous state-of-the-art methods, including LAVT~\cite{yang2022lavt} and LISA~\cite{lai2023lisa}. LAVT is composed of a vision transformer with a BERT model~\cite{kenton2019bert} for textual embedding extraction, and the whole vision branch is trainable during optimizing. LISA is based on large models, which integrates image embeddings into language models via a \textless SEG\textgreater token. As we can see from the results, we can reach competitive results with LISA, but with a much lower computational burden. There are three reasons for this phenomenon: 1) LISA uses a larger language foundation model than us(we use LLaMA-7B but LISA employs a 13B model); 2) LISA employs the SAM model~\cite{kirillov2023segment} for segmentation(SAM requires 10x times memory than CLIP during inferring); 3) the linear consumption of infusion during multimodal tuning. Some segmentation cases are shown in Figure~\ref{fig:casevisual}, illustrating the effectiveness of of proposed method.

\begin{table}[]
\centering
\caption{The results of our method on image-text classification benchmarks. The accuracy is utilized as metrics in the table.}
\label{Tab:resultCls}
\begin{tabular}{c|c|c}
\toprule 
Method & UPMC-Food101 & SNLI-VE \\
\midrule
MMBT~\cite{kiela2019supervised} & 92.10 & 74.69 \\
MaPLe~\cite{khattak2023maple} & 90.80 & 71.52 \\
PromptFuse~\cite{liang2022modular} & 82.21 & 64.53 \\
PMF~\cite{li2023efficient} & 91.51 & 71.92 \\
MiT(ours) & \textbf{93.88} & \textbf{74.71}\\
\bottomrule 
\end{tabular}
\end{table}

\subsection{Results on Image-text Classification}
On image-text classification benchmarks, we can reach the best performance on each metric, as illustrated in Table~\ref{Tab:resultCls}. Different from referring segmentation, image-text classification does not have a decoder, but a prediction head instead. We employ MaPLe~\cite{khattak2023maple} and PMF~\cite{li2023efficient} as our baselines for this task. Both MaPLe and PMF utilize existing large models while introducing new prompting strategies, achieved through concatenation or prefixing. However, MiT, being more fine-grained and delving into self-attention mechanisms, facilitates superior multimodal interactions, resulting in enhanced performance on evaluated benchmarks.

\begin{table*}[t]
\caption{The results of our method on MOSI and MOSEI datasets. \textit{Acc-2} and \textit{Acc-7} mean the accuracy of binary and seven-class classification. \textit{Acc-2} and \textit{F1} are classification metrics. \textit{MAE}, \textit{Corr}, and \textit{Acc-7} are regression metrics.}
\label{Tab:resultSen}
\centering
\begin{tabular}{c|cccc|cccc}
\toprule
\multirow{2}{*}{Models} & \multicolumn{4}{c|}{MOSI} & \multicolumn{4}{c}{MOSEI} \\
 & Acc-2/F1($\uparrow$) & MAE($\downarrow$) & Corr($\uparrow$) & Acc-7($\uparrow$) & Acc-2/F1($\uparrow$) & MAE($\downarrow$) & Corr($\uparrow$) & Acc-7($\uparrow$) \\
 \midrule 
TFN\cite{zadeh2017tensor} & 73.9/73.4 & 0.970 & 0.633 & 32.1 & 82.5/82.1 & 0.593 & 0.700 & 50.2 \\
MulT\cite{tsai2019multimodal} & 83.0/82.8 & 0.871 & 0.698 & 40.0 & 82.5/82.3 & 0.580 & 0.703 & 51.8 \\
MISA\cite{tsai2019multimodal} & 82.1/82.0 & 0.817 & 0.748 & 41.4 & 84.9/84.8 & 0.557 & 0.748 & 51.7 \\
BBFN\cite{han2021bi} & 84.3/84.3 & 0.776 & 0.775 & 45.0 & 86.2/86.1 & 0.529 & 0.767 & 54.8 \\
CubeMLP\cite{sun2022cubemlp} & 85.6/85.5 & 0.770 & 0.767 & 45.5 & 85.1/84.5 & 0.529 & 0.760 & 54.9 \\
MMIM\cite{han2021improving} & 84.1/84.0 & 0.700 & 0.800 & 46.6 & 82.2/82.6 & 0.526 & 0.772 & 54.2 \\
MiT(ours) & \textbf{86.5}/\textbf{86.5} & \textbf{0.632} & \textbf{0.858} & \textbf{49.0} & \textbf{87.2}/\textbf{87.2} & \textbf{0.509} & \textbf{0.788} & \textbf{55.2}\\
\bottomrule
\end{tabular}
\end{table*}

\subsection{Results on Sentiment Analysis}
We conduct the sentiment analysis on MOSI and MOSEI dataset. Different from image-involved tasks, sentiment analysis concentrates more on text. Therefore, the textual embeddings are directly sent to predictors without features from other modalities after the multimodal infusion. The results are shown in Table~\ref{Tab:resultSen}, our method is much better than previous methods by a large margin. One of the important reasons is that we use a large language model, while previous approaches(like MISA~\cite{hazarika2020misa} and CubeMLP~\cite{sun2022cubemlp}) generally use models such as BERT~\cite{kenton2019bert}. Despite the effectiveness of large language models, our ablation experiments can prove that MiT also plays a decisive role(Section~\ref{sec:abl}). The experiments show that LLM can not only accept common image data, but also process uncommon acoustic and facial information.

\section{Analysis}
To further study the effectiveness of MiT, we conduct some additional experiments with corresponding analysis in this section.

\begin{table}[]
\caption{The ablation study of each module and employed vision encoder in MiT. \textit{K/V-Inf} means the infusion of K and V. \textit{FF-Inf} means the infusion of feed-forward part. \textit{A.R} means the adaptive rescaling on head level. For vision encoders, \textit{P16} means the patch size is 16. The ablations are conducted on RefCOCO dataset.}
\label{Tab:ablation}
\centering
\begin{tabular}{ccc|ccc}
\toprule
\multicolumn{6}{c}{Ablation of MiT modules} \\
\midrule
K/V-Inf. & FF-Inf. & A.R. & val & testA & testB \\
\midrule
 &  &  & 51.4 & 60.1 & 50.0  \\
 & $\checkmark$ & $\checkmark$ & 69.4 & 72.1 & 65.0 \\
$\checkmark$ &  & $\checkmark$ & 72.9 & 75.6 & 67.7 \\
$\checkmark$ & $\checkmark$ &  & 72.2 & 75.4 & 67.4 \\
$\checkmark$ & $\checkmark$ & $\checkmark$ & \textbf{73.1} & \textbf{77.1} & \textbf{68.4}\\
\midrule
\multicolumn{6}{c}{Ablation of Vision Encoders} \\
\midrule
\multicolumn{3}{c|}{CLIP-Base-P16-224x} & 71.1 & 75.3 & 66.1\\
\multicolumn{3}{c|}{CLIP-Large-P14-224x} & 72.0 & 76.0 & 66.2\\
\multicolumn{3}{c|}{CLIP-Large-P14-336x} & \textbf{73.1} & \textbf{77.1} & \textbf{68.4}\\
\bottomrule
\end{tabular}
\end{table}

\subsection{Ablation Studies}\label{sec:abl}
\textbf{Modules of MiT.} There are several components in our proposed MiT, including K/V infusion(K/V-Inf.), adaptive rescaling(A.R.), feed-forward infusion(FF-Inf.), etc. The ablation study of each part on RefCOCO benchmark are shown in Table~\ref{Tab:ablation}. As we can see from the results, K/V infusion plays the most important role in MiT. When the adaptive head rescaling is removed, the performance of the model will also decrease accordingly, demonstrating the effect of this part. Besides, infusing in feed-forward module also plays a positive role for the final results.

\textbf{Selection of Image Encoders.} In our method, image encoders are crucial for segmentation, as they contains essential spatial information. Therefore, we conduct the ablation of different image backbones. As we can draw from Table6, larger backbones tends to get better performance. The performance reaches the best when we use the \textit{CLIP-Large-P14-336x} vision backbone. However, we cannot perform the ablation on text foundation model because of the hardware limitation(LLaMA-7B is the largest language model we can utilize). We believe that after replacing the larger text foundation model, the performance will be improved correspondingly.

\begin{figure}[t]
    \centering
    \begin{minipage}{1.0\linewidth}
        \centering
        \includegraphics[width=1.0\textwidth]{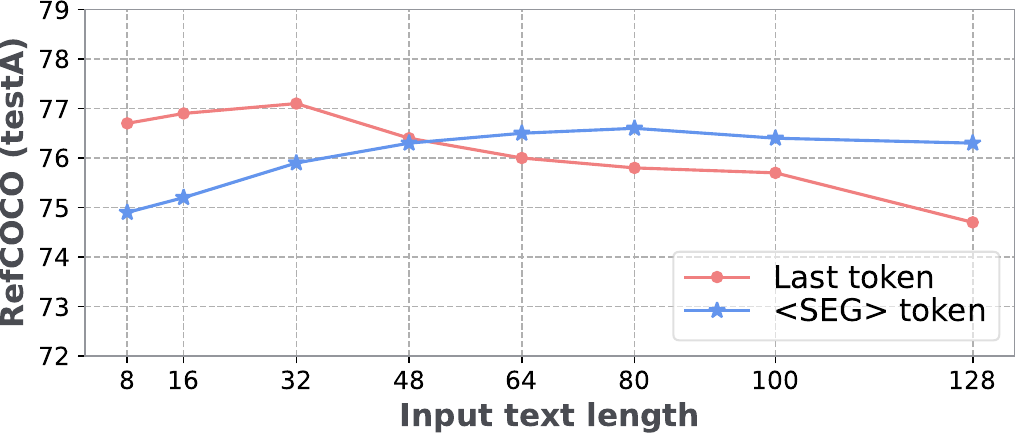}
    \end{minipage}
  \caption{The impact of text length to last-token and \textless SEG\textgreater-token schema on referring segmentation. The experiments are conducted on the testA set of RefCOCO.}
  \label{fig:textlength}
\end{figure}

\begin{figure*}[t]
    \centering
    \begin{minipage}{1.0\linewidth}
        \centering
        \includegraphics[width=0.8\textwidth]{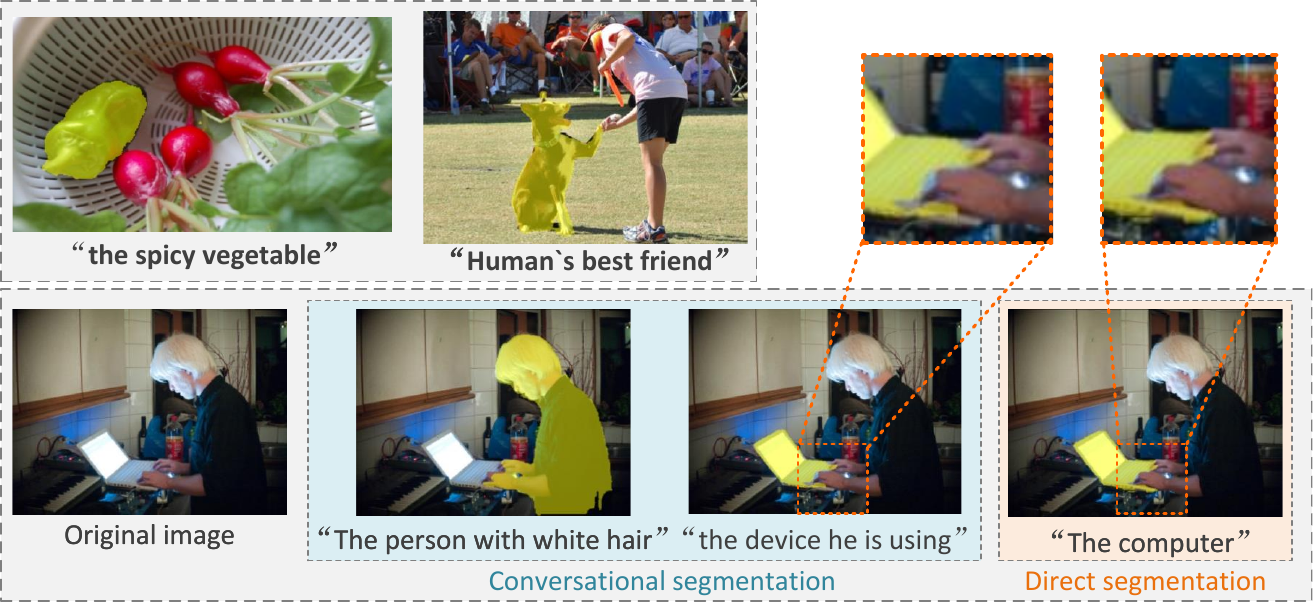}
    \end{minipage}
  \caption{The case visualization of complex reasoning(\textit{up}) and multimodal in-context understanding(\textit{down}). In complex reasoning, we describe objects more implicitly instead of identifying the object directly. For multimodal in-context understanding, we conduct the experiments in a conversation manner, which also get great segmentation results.}
  \label{fig:casevisual}
\end{figure*}

\begin{table}[]
\centering
\caption{The ablation study of infusion layers. There are 32 layers in LLaMA-7B. The experiments are conducted on RefCOCO. The performance drops when infusing in the first 1/3 layers.}
\label{Tab:layers}
\begin{tabular}{c|ccc}
\toprule
\multirow{2}{*}{Infusion layers} & \multicolumn{3}{c}{RefCOCO} \\
 & val & testA & testB \\
 \midrule
1,9,17,25,32 & 69.1 & 74.2 & 66.0 \\
9,17,25,32 & 71.9 & 75.5 & 66.8 \\
 \midrule
21,25,29,32 & 72.6 & 76.8 & 67.9 \\
17,21,25,29,32 & 72.7 & 76.7 & 68.0 \\
13,17,21,25,29,32 & \textbf{73.1} & \textbf{77.1} & \textbf{68.4}\\
\bottomrule
\end{tabular}
\end{table}

\textbf{Scales of infusion scales.} In our proposed method, we can infuse the multimodal information at arbitrary layers of LLM. There are 32 layers in LLaMA-7B in all, and the ablation results are shown in Table~\ref{Tab:layers}. We find that infusing in the last 1/3 layers of LLM achieves better results. When we infusing image information in the first 12 layers, the performance drops dramatically. The results reaches the best when we infusing the multimodal information between layer 13 and 32 with interval 4. This phenomenon can be explained by the fact that deeper layers contain richer semantic information and are thus more suitable for modality fusion.

\begin{table}[]
\caption{The ablation of involved modalities for sentiment analysis. \textit{T}, \textit{A}, and \textit{F} indicate textual, acoustic, and facial modalities, respectively. The experiments are conducted on MOSI dataset.}
\label{Tab:modalities}
\centering
\begin{tabular}{ccc|cccc}
\toprule
\multicolumn{3}{c|}{Involved modalities} & \multicolumn{4}{c}{MOSI} \\
T. & A. & F. & MAE & Corr & Acc-2/F1 & Acc-7 \\
\midrule
$\checkmark$ &  &  & 0.701 & 0.744 & 82.2/82.2 & 45.3 \\
$\checkmark$ &  & $\checkmark$ & 0.677 & 0.813 & 83.9/84.0 & 46.6 \\
$\checkmark$ & $\checkmark$ &  & 0.670 & 0.832 & 85.2/85.0 & 47.2 \\
$\checkmark$ & $\checkmark$ & $\checkmark$ & \textbf{0.632} & \textbf{0.858} & \textbf{86.5/86.5} & \textbf{49.0} \\
\bottomrule
\end{tabular}
\end{table}

\textbf{Involved modalities for sentiment analysis.} Although text dominates sentiment analysis tasks, the influence of other modalities and interaction methods cannot be ignored. The quantitative experimental results are illustrated in Table~\ref{Tab:modalities} on MOSI dataset. We find the worst performance without acoustic or facial signals, but still better than current state-of-the-art methods. When MiT is removed, meaning multimodal signals are added directly to text embedding, the performance is still behind the complete infusion method. These results prove that our performance improvement in sentiment analysis comes not only from LLM, but also from our proposed MiT.

\subsection{Difference from <TASK> Token in Multimodal Tuning}
Although MiT reaches the state-of-the art performance, it still has some difference with previous approaches. One of the main difference is the \textless TASK\textgreater token, which has been used in previous works~\cite{hazarika2020misa,ren2023pixellm}. Specifically, for segmentation, they newly introduce a learnable token \textless SEG\textgreater to the pretrained language vocabulary and use the corresponding embedding for segmentation. Instead of using an entirely new token, we utilize the last token's embedding in decoder-only LLM for different tasks.

In order to explore the difference between the two methods, we also introduce different task tokens(e.g., \textless SEG\textgreater for segmentation and \textless CLS\textgreater for classification) and conduct respective experiments. Interestingly, we find that the performance gap between them is sensitive to the length of text inputs(including the formatted prompt). The results are shown in Figure~\ref{fig:textlength}. Through the experiments, we find that when using shorter text input, the \textit{last-token} method is better; when using longer text input, the \textless SEG\textgreater-\textit{token} method performs better. Through our experiments, we observe that when utilizing shorter text inputs, the \textit{last-token} method yields superior performance, whereas with longer text inputs, the \textless SEG\textgreater-\textit{token} method performs better. Despite the variations in performance, our proposed MiT demonstrates effectiveness across different methods, highlighting its robustness. On one hand, text lengths are relatively short in the benchmarks we evaluated, and on the other hand, for simplicity, we adopt the last-token scheme for downstream tasks.

\subsection{Complex Scenario Reasoning}
In our experiments, we observe that our fine-tuned model demonstrates notable multimodal complex reasoning capabilities. Despite not receiving direct descriptions of the objects to be segmented, the model showcases the ability to segment desired parts through reasoning. For instance, as illustrated in Figure~\ref{fig:casevisual}(up), while we provide the model with the description \textit{human's best friend} instead of explicitly stating \textit{dog}, the model successfully segments the dog. The capability is inherited from pretrained LLMs, and maintained by the progressive tuning strategy. Additionally, leveraging half-precision training, the fine-tuned model achieves inference on a single NVIDIA RTX 3090 GPU in 69ms, demonstrating efficiency in both time and memory usage.

To examine the multimodal in-context understanding ability of our model, we further conduct the experiments in conversation scenarios. As shown in Figure~\ref{fig:casevisual}(down), the model can understand the contextual multimodal information(down-center). However, the performance seems worse than directly segmenting the target(down-right), which is one the the drawback of our MiT. This may be due to the fact that we did not introduce text supervision task(e.g., next-token supervision). In the future, we plan to involve the text tasks into tuning, so as to improve the ability of multimodal in-context understanding. 

\section{Conclusion}
In this paper, we proposed a new tuning method, named as multimodal infusion tuning. In our proposed MiT, we infuse the multimodal information into LLMs with a linear complexity. In addition, we also introduced an adaptive rescaling strategy to eliminate the numerical instabilities. We have conducted the experiments on seven datasets with three tasks. We reach the state-of-the-art performance but with a much lower time and memory consumption. Further analysis also reveal the multimodal understanding ability of our tuned models, including complex reasoning and multimodal in-context understanding. However, one limitation in our work is the lack of complex text understanding, which can further improve the in-context understanding capability. In the future, we are going to continue this research and involve more complex understanding tasks, including long-text reasoning and multitask tuning.

\section*{Acknowledgments}
This work was partially supported by JST SPRING to Yu Song, grant number JPMJSP2101.

\clearpage

{
\small
\bibliographystyle{ieeenat_fullname}
\bibliography{references}
}


\end{document}